\documentclass[12pt,showpacs,preprintnumbers,amsmath,amssymb,nofootinbib]{revtex4-1}
\usepackage{graphicx}
\usepackage{fancyhdr}
\usepackage{textcomp}
\usepackage{pslatex}
\usepackage{amsfonts}
\usepackage[mathcal]{eucal}
\usepackage{latexsym}
\begin{document}

\newcommand{\pderiv}[2]{\frac{\partial #1}{\partial #2}}
\newcommand{\deriv}[2]{\frac{d #1}{d #2}}

\title{Macrofotografia com um \textit{tablet}: aplica\c{c}\~oes ao ensino de Ci\^encias \newline
\small{\textit{ (Macro photography with a tablet: applications on Science Teaching)}   }  }


\vskip \baselineskip

\author{L.P. Vieira$\dag$}

\author{V.O.M. Lara$\ddag$}


\affiliation{$\dag$Instituto de F\'{\i}sica, Universidade Federal do Rio de Janeiro, \\
Rio de Janeiro, RJ, Brasil}

\affiliation{$\ddag$Instituto de F\'{\i}sica, Universidade Federal Fluminense, \\ Niter\'oi, RJ, Brasil}

\date{\today}

\begin{abstract}

Neste trabalho apresentamos uma maneira simples de obter macrofotografias (fotografias ampliadas) utilizando um \textit{tablet} ou \textit{smartphone}. Discutimos inicialmente a t\'ecnica empregada, que consiste essencialmente na coloca\c{c}\~ao de uma gota de \'agua sobre a lente da c\^amera. Em seguida, exploramos algumas aplica\c{c}\~oes ao ensino de Ci\^encias nos n\'{\i}veis fundamental e m\'edio. Conforme discutido no texto, a simplicidade e poder da t\'ecnica podem torn\'a-la uma boa ferramenta did\'atica para uso em diversas disciplinas, como Ci\^encias, Biologia e F\'isica.
\par
\textbf{Palavras-chave}: Macrofotografia, Fenomenologia, Tablet, Uso de Tecnologias no Ensino de F\'isica.

In this work we present a simple way to get Macro photography (enlarged photographs) using a tablet or phone. We initially discuss the technique, which is essentially the accommodation of a drop of water on the camera lens. Next, we explore some applications to science teaching in primary and secondary levels. As discussed in the text, the simplicity and power of the technique may make it a good teaching tool for use in various disciplines such as Science, Biology and Physics.

\end{abstract}







\maketitle

\vskip \baselineskip

\section{Introdu\c{c}\~ao}

Este trabalho tem como objetivo mostrar algumas aplica\c{c}\~oes de uma c\^amera fotogr\'afica comum encontrada em \textit{tablets} e aparelhos celulares. Com certa facilidade, usando apenas uma gota d'\'agua, podemos transformar a c\^amera em um microsc\'opio funcional e port\'atil com at\'e $150\times$ de aumento, abrindo todo um leque de aplica\c{c}\~oes para o ensino de F\'isica, Biologia e Ci\^encias no ensino fundamental e m\'edio.  A t\'ecnica n\~ao \'e nova e foi desenvolvida e utilizada anteriormente nas refer\^encias \cite{link1, link2}. Entretanto, n\~ao h\'a, at\'e onde sabemos, um trabalho que explore e divulgue aplica\c{c}\~oes desta t\'ecnica em uma perspectiva did\'atica.

\begin{figure}[!htb]
\begin{center}
\vspace{0.6cm}
\includegraphics[scale=0.50]{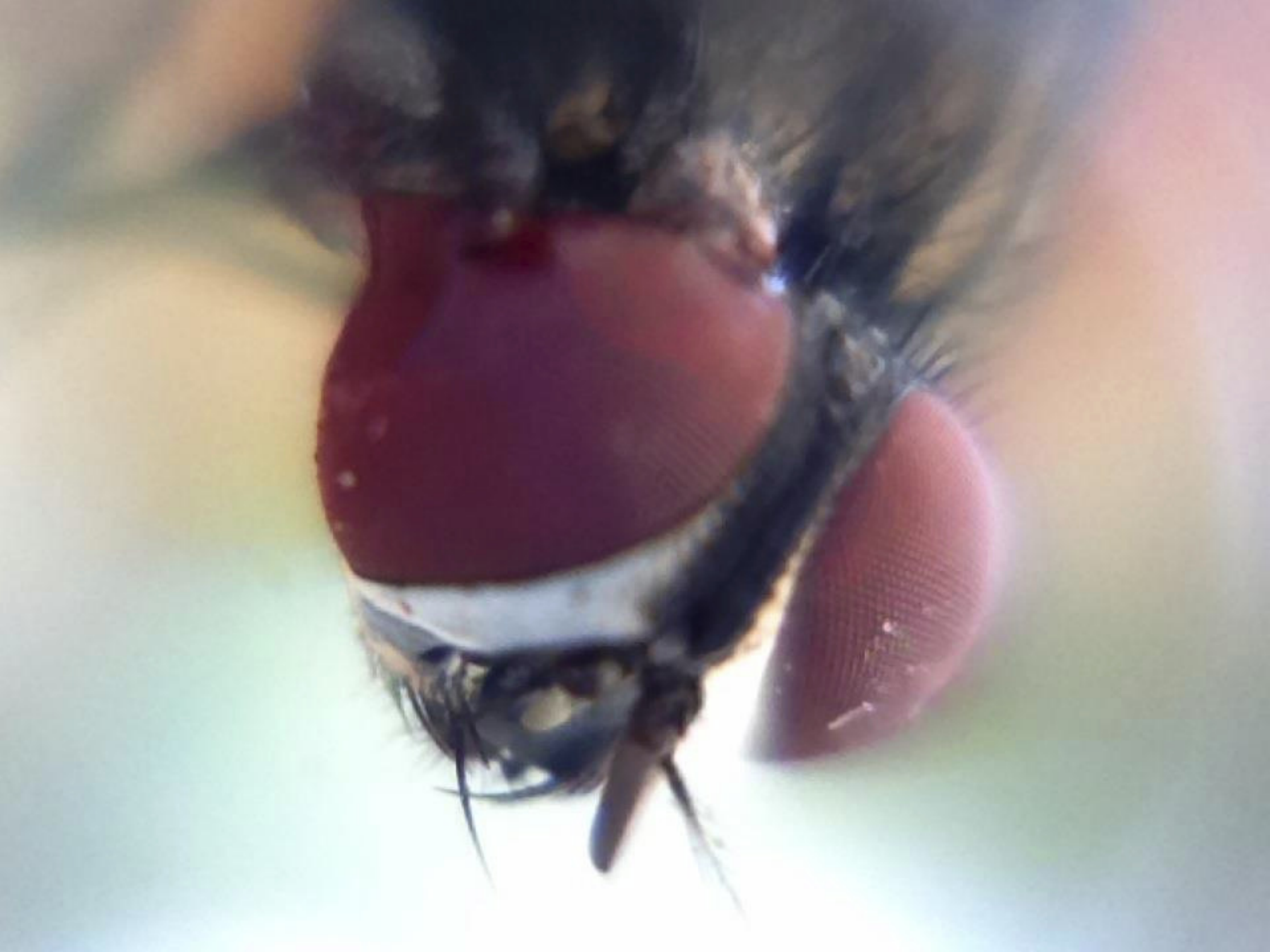}
\end{center}
\caption{Fotografia macro de uma mosca-dom\'estica (\textit{Musca domestica}) tirada com um \textit{tablet} e uma gota d'\'agua em sua lente.}
\label{figura1}
\end{figure}

Como veremos a seguir, colocando uma gota d'\'agua sobre a lente da c\^amera do aparelho obtemos fotografias como a mostrada na figura (\ref{figura1}), que podem ser exploradas em in\'umeras situa\c{c}\~oes de sala de aula. 

Dividimos este texto em se\c{c}\~oes que tratam da t\'ecnica (se\c{c}\~ao II) e das aplica\c{c}\~oes em v\'arios segmentos de ensino (se\c{c}\~ao III). E finalmente encerramos com algumas conclus\~oes e perspectivas. 

\section{A T\'ecnica}

\begin{figure}[!htb]
\begin{center}
\vspace{0.6cm}
\includegraphics[scale=0.50]{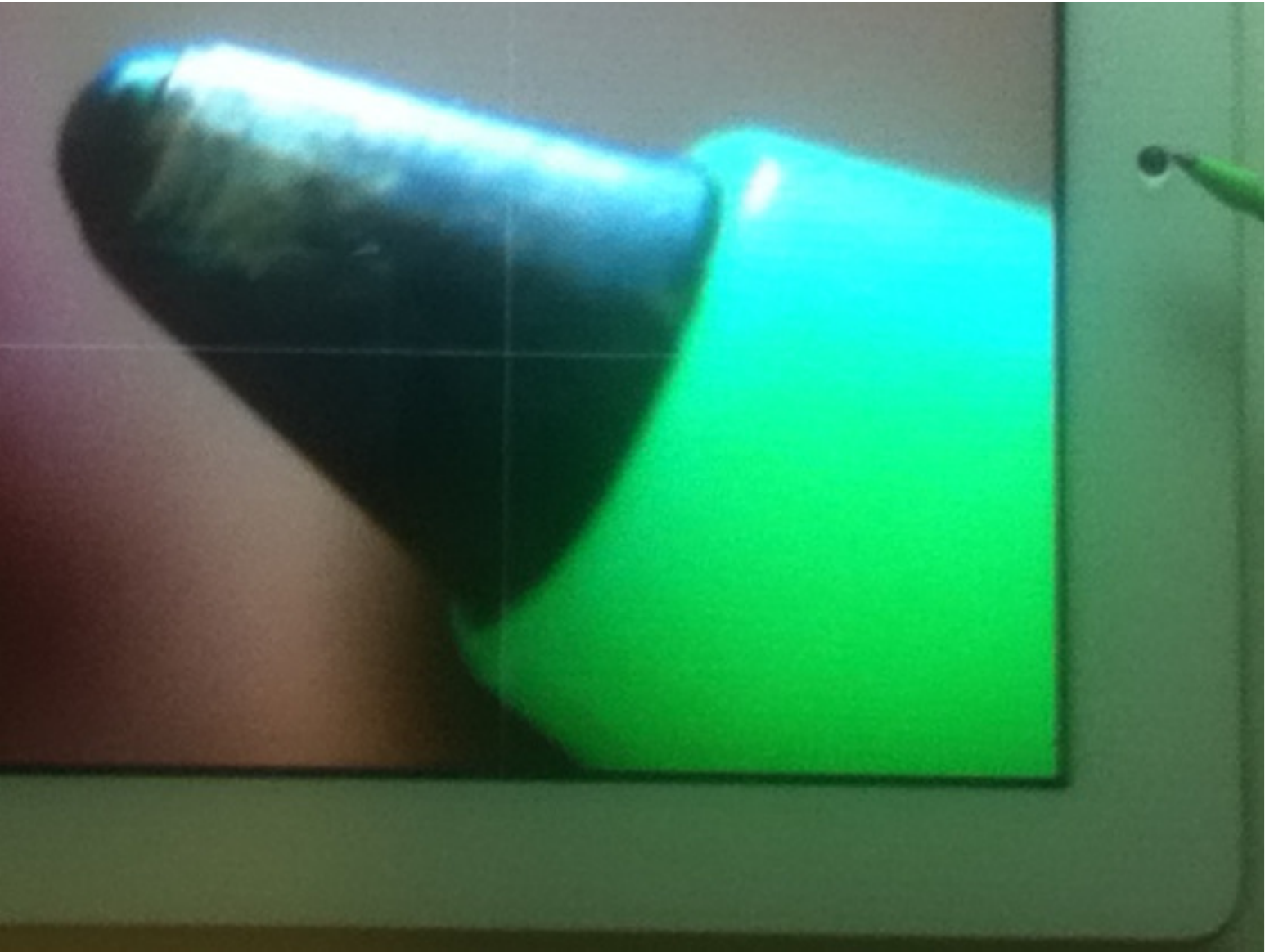}
\end{center}
\caption{Fotografia de uma caneta e sua imagem gerada em um \textit{tablet} com a t\'ecnica da gota. A caneta e a lente da c\^amera, coberta pela gota, podem ser vistas no canto superior direito da figura.Repare que para formar uma imagem n\'itida da caneta, \'e necess\'ario aproxim\'a-la da gota (cerca de 3 $mm$), conforme discutido no texto.}
\label{figura2}
\end{figure}

Para transformar uma c\^amera como a de um \textit{tablet} ou celular em um microsc\'opio funcional (e com isso obter boas fotografias macro), basta acomodar uma gota de \'agua sobre a l\^amina de vidro que protege a lente convergente da c\^amera. Com isso, efetivamente acoplamos uma segunda lente convergente cuja dioptria vai de 300 a 1000 $m^{-1}$ ($f \sim 1-3$ $mm$; obtivemos esses n\'umeros incidindo um feixe laser sobre gotas acomodadas em uma lam\'inula de microsc\'opio e medindo as dist\^ancias focais com um paqu\'imetro). Devido \`a enorme amplia\c{c}\~ao, o campo visual da c\^amera diminui consideravelmente, o que exige maior proximidade do  objeto a ser fotografado para que se possa obter uma imagem n\'itida. Para mostrar o poder de amplia\c{c}\~ao da t\'ecnica e a necessidade de aproxima\c{c}\~ao entre objeto e c\^amera, aproximamos uma caneta verde da mesma, conforme pode-se ver na  figura~(\ref{figura2}).

\begin{figure}[!htb]
\begin{center}
\vspace{0.6cm}
\includegraphics[scale=0.50]{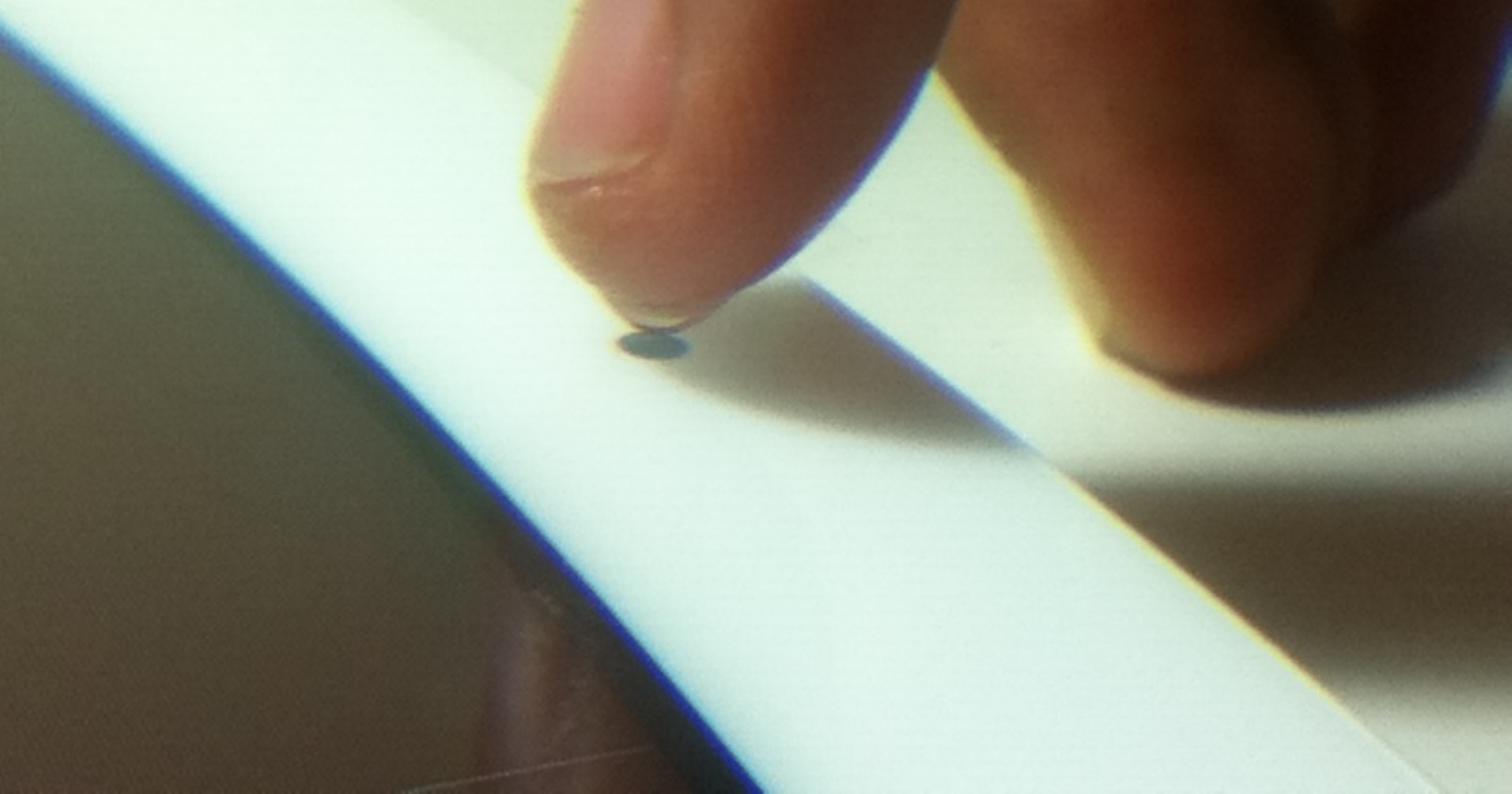}
\end{center}
\caption{Fotografia de uma gota sendo acomodada \`a l\^amina protetora da c\^amera de um \textit{tablet}.}
\label{figura3}
\end{figure}

Para aplicar a gota sobre o dispositivo, basta molhar um de seus dedos e mov\^e-lo cuidadosamente (para que a gota permane\c{c}a em seu dedo) at\'e a l\^amina. Ao encostar a por\c{c}\~ao d'\'agua que est\'a em seu dedo na l\^amina, as for\c{c}as de ades\~ao e coes\~ao \cite{moises} dar\~ao conta de acomod\'a-la, como se v\^e na figura (\ref{figura3}). 

\vspace{1 cm}

\'E comum em muitos aparelhos a exist\^encia de uma segunda c\^amera localizada na frente do dispositivo, voc\^e \'e livre para escolher onde acomodar a gota. Entretanto, vale \`a pena consultar o manual do dispositivo, pois essa c\^amera tende a ter qualidade inferior, ou seja, produz imagens com menor resolu\c{c}\~ao.
Frequentemente a luminosidade ambiente n\~ao \'e suficiente para captura de boas fotos (isso se deve a diminui\c{c}\~ao do campo visual da c\^amera). Para contornar essa dificuldade podemos usar uma lanterna para iluminar o objeto de estudo. 

Uma vez descrito o procedimento e condi\c{c}\~oes necess\'arias para a produ\c{c}\~ao de fotografias macro discutidas nesta se\c{c}\~ao, vamos passar a algumas aplica\c{c}\~oes did\'aticas desta t\'ecnica.

\section{Aplica\c{c}\~oes}

\subsection{F\'isica}

A t\'ecnica descrita acima foi utilizada nas aulas de F\'isica de alguns col\'egios de n\'ivel m\'edio onde um dos autores leciona. Foram exploradas duas perspectivas distintas desse arranjo. A primeira delas teve como objetivo motivar o ensino das lentes e a segunda explorou a bi-refring\^encia de gr\~aos de a\c{c}\'ucar. 

\begin{figure}[!htb]
\begin{center}
\vspace{0.6cm}
\includegraphics[width=7cm]{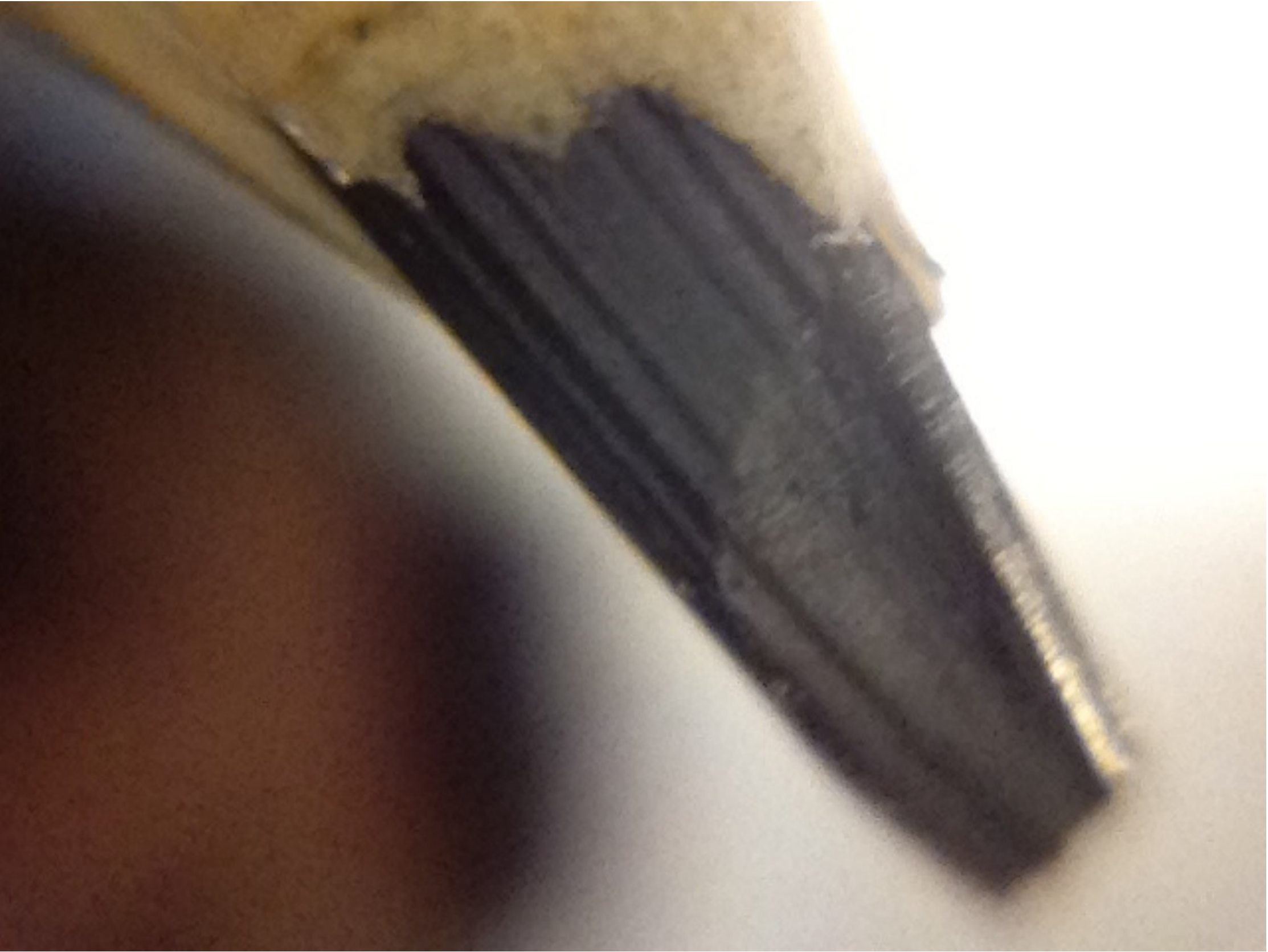}
\includegraphics[width=7cm]{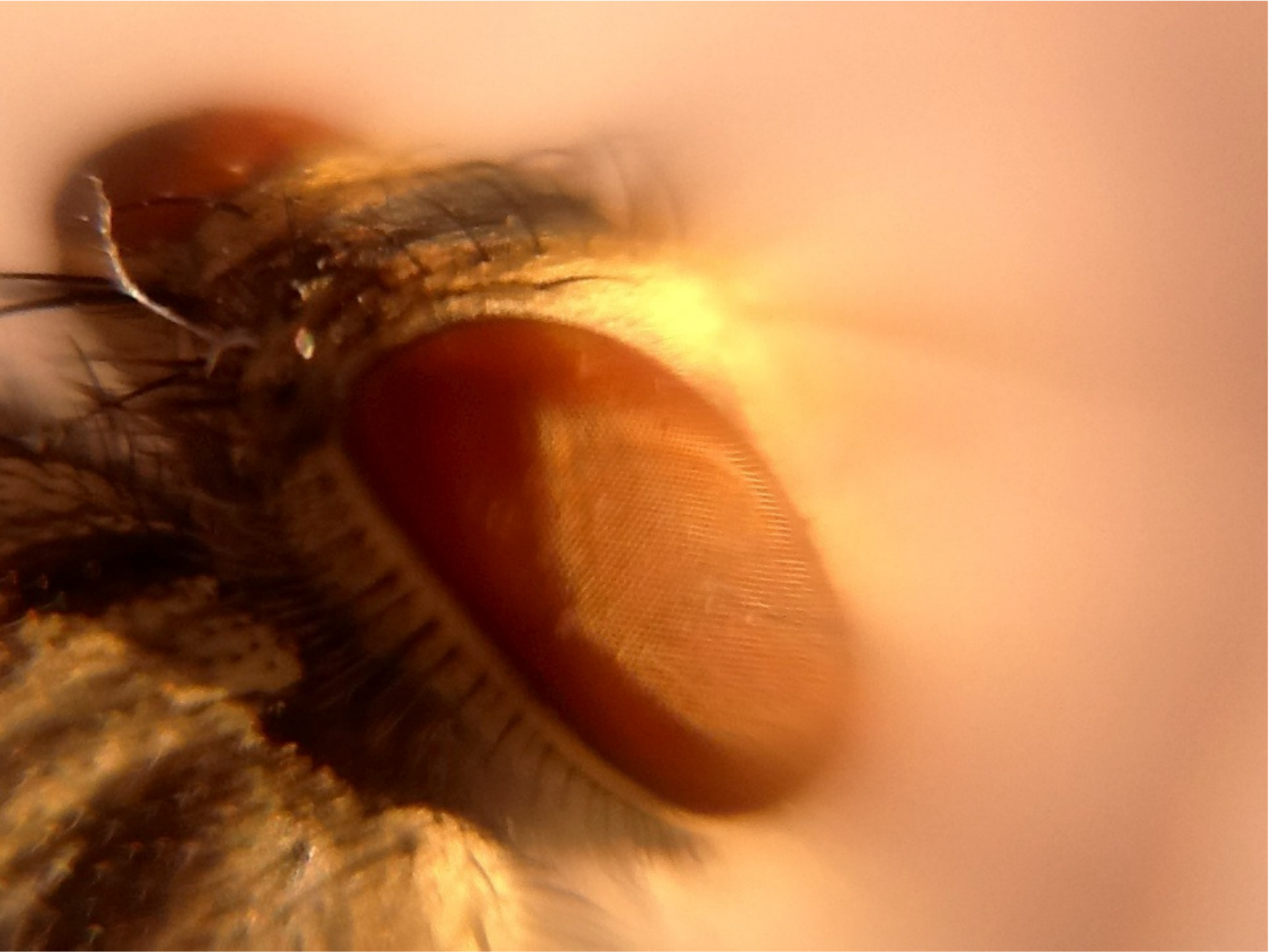}
\end{center}
\caption{\`A esquerda temos um l\'apis ampliado (cerca de $60\times$) e \`a direita uma mosca varejeira (ou mosca-da-carne), pertencente \`a fam\'ilia Sarcophagidae.}
\label{figuras4_5}
\end{figure}

No estudo das lentes, a t\'ecnica foi apresentada aos educandos do $1^{\circ}$ ano, com o objetivo de motiv\'a-los para o assunto em quest\~ao. Pedimos que os alunos aplicassem a t\'ecnica para obter fotografias de 3 objetos que eles julgassem interessantes para uma observa\c{c}\~ao ampliada. Recebemos imagens de diversas esp\'ecies de animais, vegetais e objetos de uso cotidiano. Alguns exemplos podem ser vistos na figura (\ref{figuras4_5}).

\begin{figure}[!htb]
\begin{center}
\vspace{0.6cm}
\includegraphics[width=7cm]{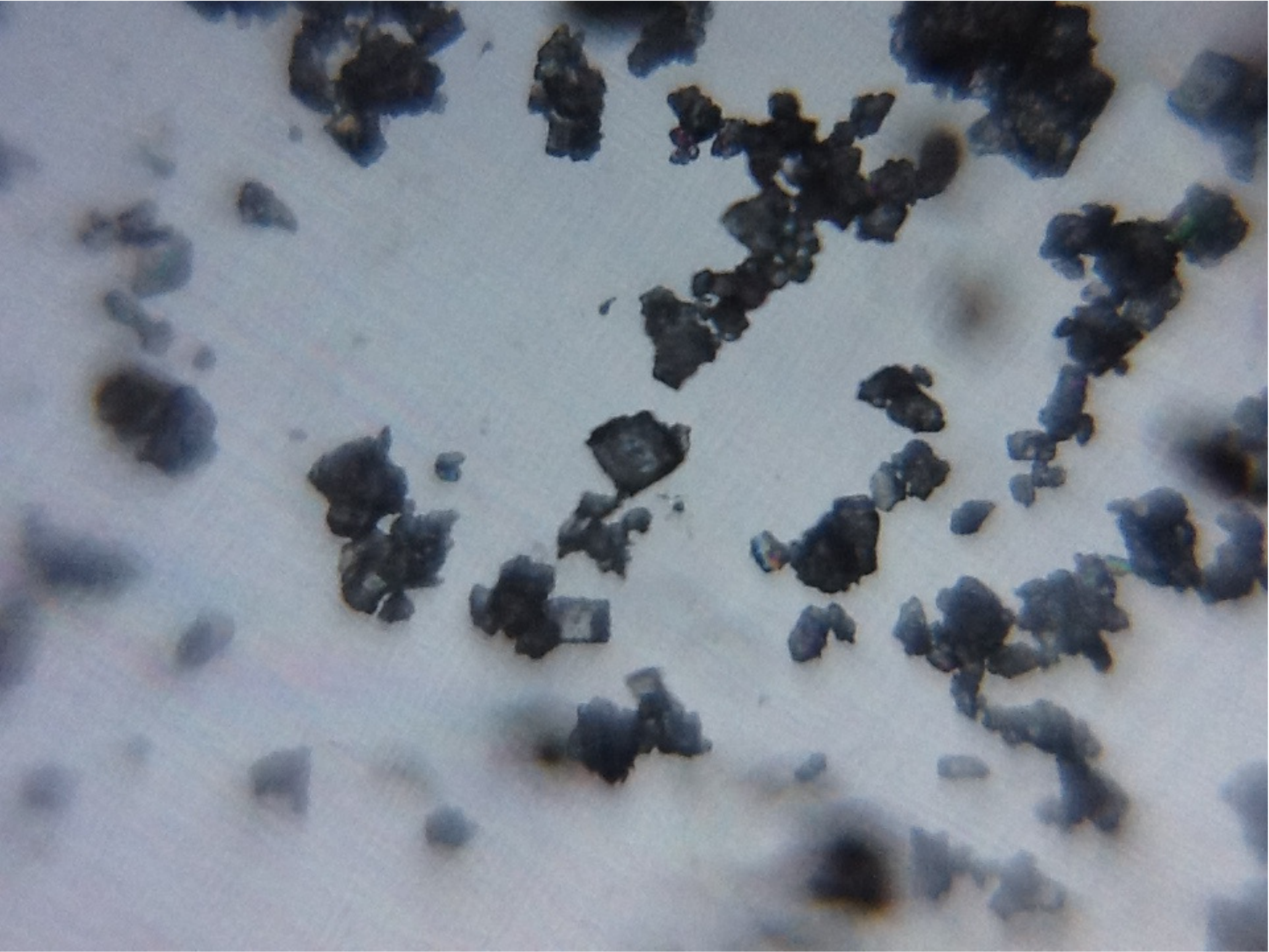}
\end{center}
\caption{Imagem ampliada de gr\~aos de a\c{c}\'ucar, sobre um polar\'oide, evidenciando o fen\^omeno da bi-refring\^encia.}
\label{figura6}
\end{figure}

A segunda aplica\c{c}\~ao, referente aos gr\~aos de a\c{c}\'ucar, consistiu em repousar aleatoriamente gr\~aos de a\c{c}\'ucar sobre uma l\^amina polarizadora \cite{Hecht}, iluminada por baixo por uma fonte de luz difusa. Ao aproximar a c\^amera desse arranjo podemos perceber g\~aos iluminados (de maneira transl\'ucida) e totalmente apagados [veja a figura (\ref{figura6})]. 

\'E importante notar o papel da macrofotografia neste estudo: s\'o podemos observar essas caracter\'isticas com uma boa amplia\c{c}\~ao da imagem. Os gr\~aos t\^em em m\'edia um di\^ametro de $0.1$ $mm$. Devido a esta pequena escala de comprimento, n\~ao somos capazes de observar a luz que atravessa os gr\~aos a olho nu. \`A medida que a luz plano-polarizada atravessa um determinado gr\~ao, ela pode seguir em duas dire\c{c}\~oes distintas; entretanto, apenas a que segue em dire\c{c}\~ao \`a c\^amera pode ser observada. Devido a isto, temos algumas imagens distintas, de gr\~aos transl\'ucidos e de gr\~aos mais escuros\cite{Hecht}.

\subsection{Ci\^encias no Ensino Fundamental}

No ensino fundamental exploramos a t\'ecnica com alunos do $6^{\circ}$ ao $9^{\circ}$ anos em v\'arias frentes distintas, como o estudo dos solos e gr\~aos, e de partes do corpo humano. Outras aplica\c{c}\~oes envolveram o estudo da Bot\^anica, da Entomologia e da vis\~ao em cores.

No estudo dos solos, um dos autores deste trabalho utilizou a t\'ecnica para que os alunos do $7^{\circ}$ ano pudessem perceber as diferentes caracter\'isticas que minerais podem apresentar. Como exemplo, foram feitas compara\c{c}\~oes entre pedras magm\'aticas e calc\'arias. E assim, depois de motivar os estudantes, discutimos a origem de cada uma delas. 

J\'a com o corpo humano, partes como a l\'ingua foram exploradas a fim de se estudar as formas das papilas gustativas. No caso das unhas e dos dedos foram evidenciadas min\'usculas ranhuras e imperfei\c{c}\~oes. Al\'em disso, p\^ode-se discutir a necessidade da manuten\c{c}\~ao da higiene pessoal, uma vez que diversas part\'iculas de poeira foram observadas. P\^ode-se constatar tamb\'em a exist\^encia de p\^elos entre os sulcos digitais humanos.

\begin{figure}[!htb]
\vspace{0.6cm}
\includegraphics[scale=0.40]{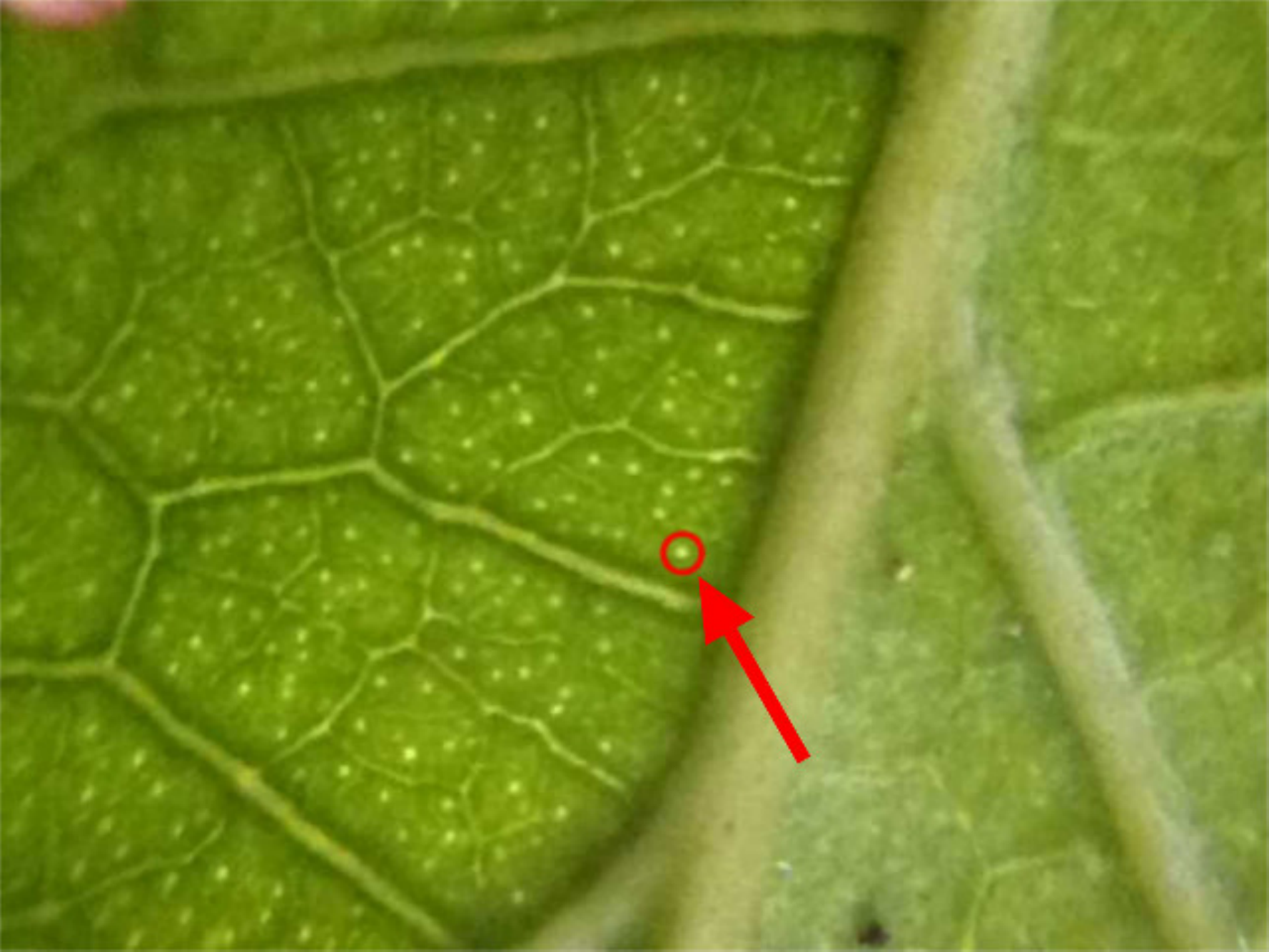}
\includegraphics[scale=0.40]{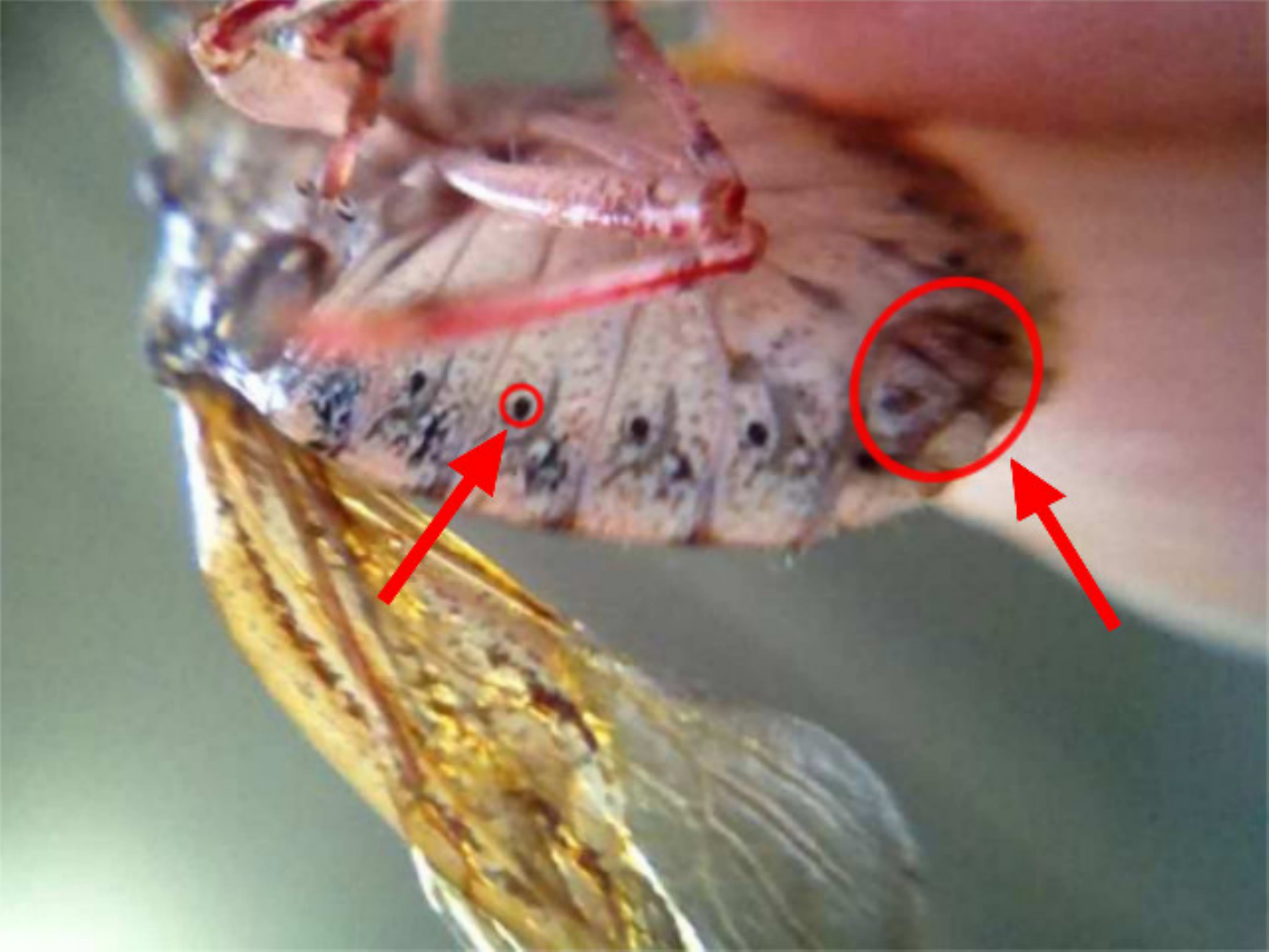}
\caption{\`A esquerda, temos a imagem de uma folha de uma planta do g\^enero \textit{Malvaviscus}, fam\'ilia Malvaceae, a mesma do algodoeiro, onde destacamos os est\^omatos, aberturas destinadas \`a trocas gasosas da fotoss\'intese/respira\c{c}\~ao/transpira\c{c}\~ao. J\'a o inseto \`a direita \'e um percevejo f\^emea, da ordem Hemiptera, fam\'{\i}lia Pentatomidae. Na foto podemos ver os espir\'aculos, aberturas respirat\'orias em cada segmento do abdome, e no final do abdome est\~ao as valvas do ovipositor, que s\~ao parte da genit\'alia feminina externa.}
\label{figuras7_8}
\end{figure}

No estudo da Bot\^anica e da Entomologia, um professor de Biologia coletou diversos esp\'ecimes de vegetais e insetos. Com base nas macrofografias dos esp\'ecimes discutiu-se com alguns estudantes diversas caracter\'{\i}sticas morfol\'ogicas destes seres. Na figura \`a esquerda (\ref{figuras7_8}), podemos ver uma fotografia ampliada de uma folha. Nesta imagem, o ponto branco destacado refere-se \`a c\'elulas macrosc\'opicas, conhecidas como est\^omatos. Tamb\'em na figura (\ref{figuras7_8}), temos a fotografia de um percevejo f\^emea. Real\c{c}amos nesta imagem parte do aparelho respirat\'orio e sexual desta esp\'ecie.

\begin{figure}[!htb]
\vspace{0.6cm}
\includegraphics[scale=0.30]{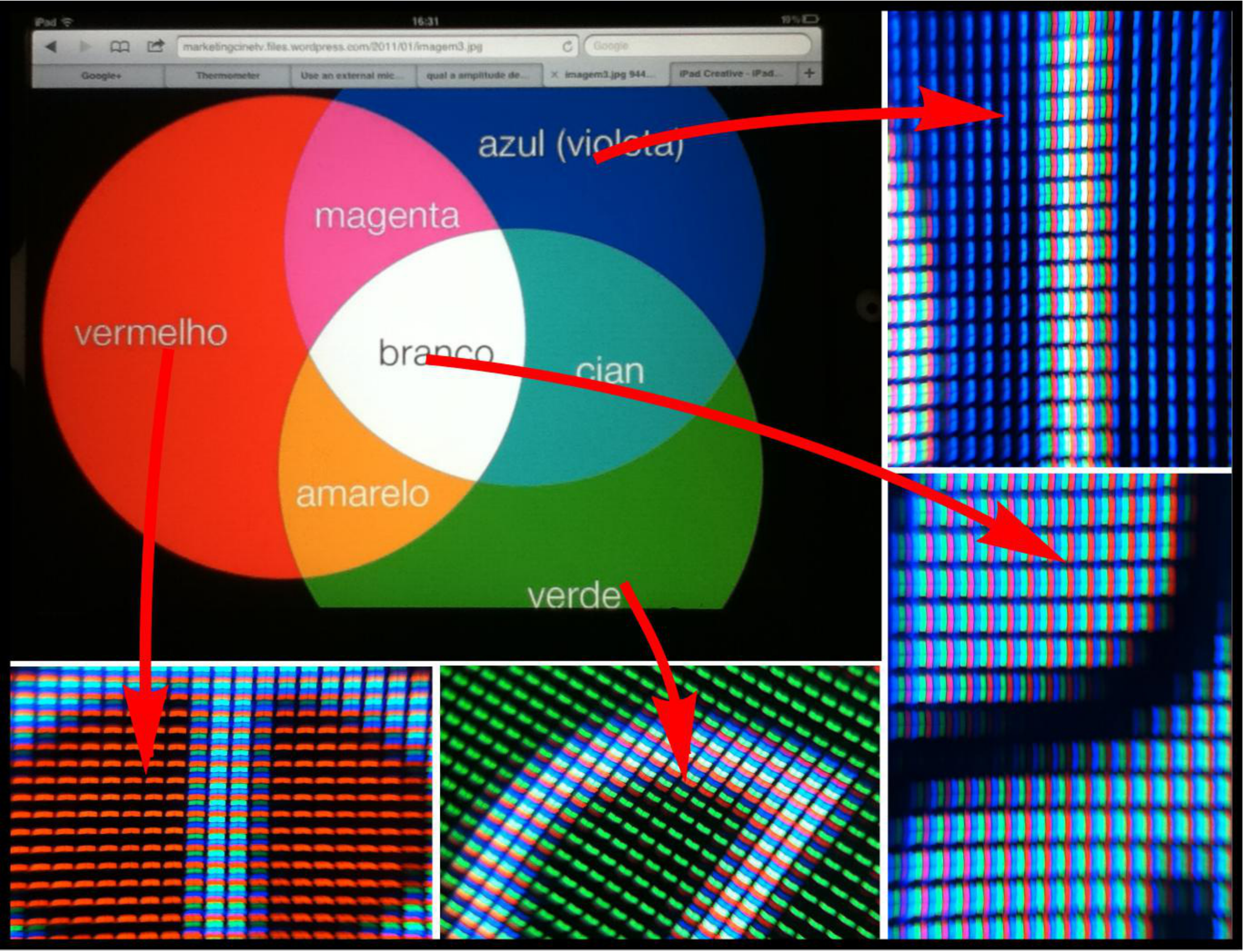}
\caption{Imagem que mostra como a combina\c{c}\~ao das cores fundamentais (vermelho, verde e azul) gera todas as cores que vemos em uma tela LCD (em cores na vers\~ao \textit{on-line}.)}
\label{figura9}
\end{figure}

Tamb\'em aplicamos a t\'ecnica da gota para ampliar um \textit{display} LCD de um \textit{tablet} de um dos estudantes do $9^{\circ}$ ano do Ensino Fundamental. Nosso objetivo era estudar a forma\c{c}\~ao das cores, por meio da mistura de comprimentos de ondas distintos. Na figura (\ref{figura9}) temos a foto em tamanho real de uma tela LCD de um \textit{tablet}. Em volta desta imagem, podemos observar as imagens ampliadas de cada uma das regi\~oes indicadas. Podemos ver que um \textit{pixel} \'e formado por 3 c\'elulas distintas que deixam passar, respectivamente, vermelho, verde e azul, as chamadas cores prim\'arias. Controlando a intensidade da luz de cada uma dessas c\'elulas, o LCD forma um ponto com uma cor secund\'aria espec\'{\i}fica. Como cada \textit{pixel} tem por volta de 0,096 mm, n\~ao somos capazes de distingui-los a olhu nu, o que nos d\'a a impress\~ao de uma imagem cont\'inua. Na imagem (\ref{figura9}) podemos ver que o branco se forma quando os \textit{pixels} deixam passar com a mesma intensidade o verde, o vermelho e o azul. Todas as outras cores s\~ao formadas por este mecanismo: basta ajustar a intensidade de cada uma das cores prim\'arias.

\subsection{Medidas com macrofotografias}

\begin{figure}[!htb]
\begin{center}
\includegraphics[scale=0.60]{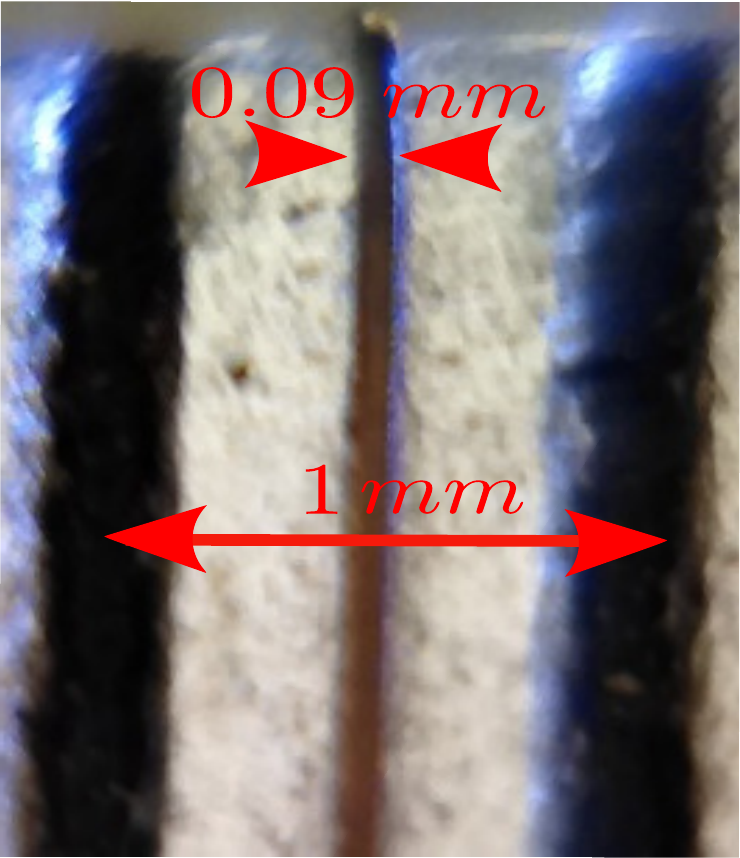}
\end{center}
\caption{Imagem de um fio cabelo preso a um paqu\'imetro. Conforme discutido no texto estimamos a espessura deste em 0.09 $mm$.}
\label{figura10}
\end{figure}

Trabalhamos tamb\'em no\c{c}\~oes de ordens de grandeza a partir das macrofotografias. Medimos, por exemplo, a espessura de um fio de cabelo. A ideia foi a seguinte: prendemos o fio de cabelo sobre o n\^onio de um paqu\'imetro, utilizando fita adesiva. Em seguida, tiramos uma fotografia ampliada desta montagem, conforme pode-se ver na figura (\ref{figura10}).

Depois, com o aux\'ilio de um programa de edi\c{c}\~ao de imagens (nesse estudo utilizamos o software livre \textit{Gimp} \cite{gimp}), medimos o comprimento de refer\^encia dado pelo n\^onio (um mil\'imetro) e o do fio de cabelo, contando os \textit{pixels}. Com este procedimento, fomos capazes de estimar o di\^ametro do fio de cabelo em $0.09$ $mm$, relativamente pr\'oximo do que foi obtido  por meio de difra\c{c}\~ao na refer\^encia \cite{fio_de_cabelo}.

Al\'em disto, tamb\'em estimamos com os estudantes o aumento obtido na fotografia da figura (\ref{figura10}). Para fazer isto, imprimimos em papel a imagem em v\'arios tamanhos distintos, at\'e o limite de boa qualidade da impress\~ao (fizemos isto para que pud\'essemos determinar o aumento relativo - e posteriormente absoluto - das dimens\~oes). Com uma r\'egua medimos o tamanho efetivo do cabelo nas folhas. De posse destes dados, os alunos tamb\'em foram capazes de estimar a ampli\c{c}\~ao da imagem impressa no papel.   

\section{Conclus\~oes e Perspectivas}

Neste trabalho discutimos diversas aplica\c{c}\~oes  de macrofotografias obtidas com a t\'ecnica da gota nos n\'iveis fundamental e m\'edio. Entretanto, n\~ao pretendemos ter feito aqui uma listagem exaustiva de todas as poss\'iveis aplica\c{c}\~oes. O leitor pode realizar suas pr\'oprias experi\^encias e prospostas de atividades. Finalmente, cabe ressaltar a simplicidade e a riqueza de possibilidades oferecidas por esta t\'ecnica, que pode  se tornar uma boa ferramenta did\'atica em sala de aula.

\medskip
 Os autores s\~ao gratos \`a CAPES pelo apoio financeiro, e ao IFRJ-S\~ao Gon\c{c}alo e Centro Educacional de Niter\'oi pela permiss\~ao para a realiza\c{c}\~ao das atividades que relatamos neste trabalho. Tamb\'em somos especialmente gratos ao professor  Leonardo Rocha (IFRJ) por fornecer diversas informa\c{c}\~oes relativas aos esp\'ecimes coletados e participa\c{c}\~ao nas atividades relacionadas \`a Biologia, ao professor  Carlos E. Aguiar (IF-UFRJ) e ao doutorando em F\'isica  Marlon F. Ramos (PUC-RJ) pela leitura criteriosa e sugest\~oes interessantes, e a todos os estudantes envolvidos nestas atividades.




\vskip 2\baselineskip

\vskip 2\baselineskip

\end{document}